\documentclass[aps,prx,twocolumn,superscriptaddress,nofootinbib]{revtex4-2}
\usepackage{graphics}
\usepackage{amsmath}
\usepackage{physics}
\usepackage{graphicx}
\usepackage{color}
\usepackage{amsfonts,wasysym}
\usepackage{amssymb}
\usepackage{float,tikz}

\newcommand{\bs}[1]{\boldsymbol{#1}}
\newcommand{\pauli}[1]{\mathtt{#1}}

\DeclareMathOperator*{\argmax}{arg\,max}

\newcommand{\prlsection}[1]{{\it #1}:--}

\begin{document}

\title{Learning to Classify Quantum Phases of Matter with a Few Measurements} 

\author{Mehran Khosrojerdi}
\affiliation{Department of Physics and Astronomy, University of Florence,
via G. Sansone 1, I-50019 Sesto Fiorentino (FI), Italy}

\author{Jason L. Pereira} 
\affiliation{ INFN Sezione di Firenze, via G. Sansone 1, I-50019, Sesto Fiorentino (FI), Italy }

\author{Alessandro Cuccoli}
\affiliation{Department of Physics and Astronomy, University of Florence,
via G. Sansone 1, I-50019 Sesto Fiorentino (FI), Italy}
\affiliation{ INFN Sezione di Firenze, via G. Sansone 1, I-50019, Sesto Fiorentino (FI), Italy }

\author{Leonardo Banchi}\email{leonardo.banchi@unifi.it}
\affiliation{Department of Physics and Astronomy, University of Florence,
via G. Sansone 1, I-50019 Sesto Fiorentino (FI), Italy}
\affiliation{ INFN Sezione di Firenze, via G. Sansone 1, I-50019, Sesto Fiorentino (FI), Italy }

\date{\today}

\begin{abstract}
	We study the identification of quantum phases of matter, at zero temperature, when only part of 
	the phase diagram is known in advance. Following a supervised learning approach, 
	we show how to use our previous knowledge to construct an observable capable of classifying the 
	phase even in the unknown region. 
	By using a combination of classical and quantum techniques, such as tensor networks, kernel 
	methods, generalization bounds, quantum algorithms, and shadow estimators, we show that, in some cases, 
	the certification of new ground states can be obtained with a polynomial number of measurements. 
	An important application of our findings is the classification of the phases of matter obtained in quantum 
	simulators, e.g.,~cold atom experiments, capable of efficiently 
	preparing ground states of complex many-particle systems
	and applying simple measurements, e.g.,~single qubit measurements, 
	but unable to perform a universal set of gates. 
\end{abstract}

\maketitle

\prlsection{Introduction} 
Identifying phases of quantum matter is a complex task \cite{sachdev1999quantum,zinn2007phase} that 
requires the extraction of possibly non-local features from the ground states of quantum many-particle systems. 
Given the recent remarkable success of machine learning methods to extract general laws from a 
few known examples \cite{carleo2019machine}, many studies applied machine learning techniques 
to learn quantum phases of matter 
\cite{cong2019quantum,cea2024exploring,monaco2023quantum,huang2022provably,dong2019machine,uvarov2020machine,carrasquilla2017machine,li2024ensemble} -- see \cite{carleo2019machine} for more references. 
Among the different learning methods that can be applied to ``quantum data''  \cite{schuld2021supervised},
such as the ground states of quantum many-particle systems, 
kernel methods stand out for their interpretability, since learning consists in finding 
a decision hyperplane in a Hilbert space \cite{cristianini2000introduction,schuld2021supervised,montalbano2024quantum} -- see Appendix \ref{sec:svm} for a brief introduction. 
The problem of classifying quantum phases of matter with kernel methods has been recently considered in Refs.~\cite{sancho2022quantum,wu2023quantum}
 -- see Appendix \ref{sec:comparison} for a detailed comparison between the different approaches.

In a quantum phase classification problem, the state we want to classify is the ground state of a parametrised Hamiltonian with a specific form. For some parameter values we may have full (classical) knowledge of the phases. 
In many physically relevant scenarios, this comes from our ability to classically simulate the true 
ground state of the Hamiltonian, e.g.,~using matrix product states (MPS) 
or other tensor network methods \cite{schollwock2011density,banuls2023tensor}. 
This suggests a machine learning approach where training is purely classical, 
performed on states within the known parameter region, with the ultimate aim of learning
a general rule (e.g., an observable or a measurement) that can be applied to classify ground states 
unseen during training.

One physically motivated reason to \emph{test} a machine learning model with different ground states 
may be that classical approximation methods are not accurate in that parameter regime. 
A central result of our analysis is that, 
if such ground state may be prepared in a quantum device such as
quantum simulators \cite{altman2021quantum}, e.g.,~by cooling a quantum many-body system, 
the classification of new ground states, unseen during training, can be obtained with 
single-qubit measurements. 
This result is derived by pairing kernel methods, tensor network techniques,
generalization bounds \cite{banchi2021generalization,banchi2024statistical}, and 
the certification strategy from Ref.~\cite{huang2024certifying}.
Our algorithm is hybrid, since training is performed classically, 
while testing new states can be done in different ways, e.g.,~in quantum devices. 
Since single-qubit measurements are routinely performed in quantum simulators, 
an universal quantum computer is never required. 
Nonetheless, the same approach can also be applied to different situations, e.g.,~when 
we have classical knowledge of the test ground states or when they are approximated 
in a quantum computer via parametric quantum circuits 
\cite{bharti2022noisy}.

\prlsection{Problem description}
We are interested in getting an approximation of the phase diagram for the ground state of a quantum
many-particle Hamiltonian $H(\bs x)$ that depends on classical external
parameters $\bs x=(x_1,x_2,\dots)$. We focus on settings satisfying two main
assumptions: i) the ground state $\ket{\psi(\bs x)}$ of $H(\bs x)$ can be
obtained with sufficiently high precision, for certain values of $\bs x$, using
classical methods based on tensor networks \cite{banuls2023tensor}; ii) within
the parameter subspace where the above methods are accurate, there is a
(usually smaller) subspace $\mathcal S$ where the phase of $\ket{\psi(\bs x)}$
is known for all $\bs x \in \mathcal S$. The task is to use our knowledge from
subregion $\mathcal S$ to construct an observable $\Lambda$ that allows us to
predict the phase of a ground state. If we are given copies of a state
$\ket{\psi(\bs x)}$ for unknown $\bs x \in \mathcal S$ (e.g., as the outcome of
an experiment), we can evaluate the expectation value of $\Lambda$ to determine
its phase (in-distribution generalization). Alternatively, for a test point
outside the known subregion ($\bs x\notin \mathcal S$), we can use $\Lambda$ to
learn the phase diagram outside $\mathcal S$ (out-of-distribution generalization) 
with the help of a quantum computer or simulator, e.g.,~a cold-atom experiment, able 
to efficiently find the ground state when $\bs x\notin \mathcal S$.

To construct $\Lambda$ 
we use a supervised learning approach, where we first build a training set $\{\ket{\psi(\bs x_n)},y_n\}$, 
for different choices of $\bs x_n$ from $\mathcal S$, with 
our best approximation of the true ground state  of $H(\bs x_n)$, $\ket{\psi(\bs x_n)}$, 
and its known phase $y_n$. 
The index $n$ is a discrete index labelling the $M$  training pairs. 
To simplify the presentation, we use the same notation
$\ket{\psi(\bs x)}$ for the true ground state and our best approximation of it. 
In choosing the training algorithm we considered 
two desirable properties: i) it should
work efficiently with both
classical (e.g.,~tensor network)
and shallow quantum circuit approximations  
of many-body quantum states whose Hilbert space is 
exponentially large, and ii) it should give, as an output, a quantity that can be efficiently 
measured on a quantum device. Because of those requirements, we decided to focus on kernel methods 
\cite{schuld2021supervised}.

For any two points $\bs x_n$ 
and $\bs x_m$, the kernel is defined as 
\begin{align}
	k(\bs x_n,\bs x_m) &= \Tr[\rho(\bs x_n)\rho(\bs x_m)] = |\bra{\psi(\bs x_n)}\psi(\bs x_m)\rangle|^2,
	\label{eq:kernel}
\end{align}
where $\rho(\bs x) = \ket{\psi(\bs x)}\!\!\bra{\psi(\bs x)}$.
Both training and testing in kernel methods only require the above kernel matrix, and never use
the full ground states, which live in a exponentially large Hilbert space. As such, it can be 
easily extended to settings involving many-particle systems, provided that the inner 
product \eqref{eq:kernel} can be efficiently computed, as with tensor networks, 
shallow quantum circuits, or a combination of both \cite{cramer2010efficient,rudolph2023synergistic,huang2024certifying}. 
Another, more physical, reason to focus on kernel methods for classifying phases of matter is 
that, for nearby points $k(\bs x,\bs x+\dd\bs x)$, the kernel reduces to the 
fidelity susceptibility, which has been extensively used to predict phase transitions 
\cite{zanardi2006ground,campos2007quantum,banchi2014quantum}.
Therefore, it is reasonable to expect that the kernel entries \eqref{eq:kernel} contain 
all the information about the quantum phases.

In kernel methods, training consists in finding a decision function $y=f(\bs x)$ that can infer the label $y$ (namely the quantum phase) given a certain input $\bs x$. 
The training algorithm involves a convex optimization routine, 
which always reaches the global optimum, and whose outcome is a weight vector $\bs\alpha$ of 
real numbers, together with the baseline $b \in\mathbb R$. 
From $\bs\alpha$ and $b$, the algorithm constructs the {\it decision function} as a ``kernel expansion of the data''
$\sum_n \alpha_n y_n k(\bs x_n,\bs x)+b$, where $\bs x_n$ are the points used during training, and 
$\bs x$ is a new point, possibly not belonging to $\mathcal S$, that we want to test. 
The crucial observation, already noted in \cite{schuld2021supervised}, is that the kernel 
expansion can also be expressed as the expectation value of of an observable $\Lambda$ over 
the state ${\rho(\bs x)}$. 
For instance, in the simplest case of  binary classification, with two phases labeled as $y=\pm1$, 
the model prediction is expressed as 
\begin{align}
	y_{\rm predicted} &= {\rm sign}\left(\sum_n y_n\alpha_n k(\bs x_n,\bs x)+b\right) =\\ 
										&= {\rm sign}\left(\bra{\psi(\bs x)}\Lambda \ket{\psi(\bs x)}+b\right),
	\label{eq:y predicted}
\end{align}
where in the second line we have defined the observable 
\begin{equation}
	\Lambda = \sum_n y_n\alpha_n \rho(\bs x_n), 
	\label{eq:order parameter}
\end{equation}
as a weighted combination of density matrices. In other words, kernel methods allow for the construction 
of a {\it decision operator} $\Lambda$ that, being Hermitian, can be measured in a quantum device capable 
of preparing the state $\ket{\psi(\bs x)}$ even for those values of $\bs x$ for which the true phase is 
unknown. 
Decision functions with multiple labels are mapped into multiple binary decisions, each with a 
different decision observable. 
For numerical results, as a classical training algorithm, we focus on Support 
Vector Machines (SVMs) \cite{cristianini2000introduction}, which can be computed efficiently 
using available numerical libraries \cite{chang2011libsvm}. See Appendix~\ref{sec:svm} for 
a self-contained introduction.

We now discuss the evaluation of the kernel entries \eqref{eq:kernel}. For training and 
in-distribution generalization, or in any other case when the true states are 
replaced by their tensor network approximations, inner product \eqref{eq:kernel} can be computed 
efficiently on classical hardware \cite{schollwock2011density}. 
When the true states are approximated via a shallow quantum circuit, it can be 
computed using different methods such as the swap test or inversion test \cite{schuld2021supervised}.
For out-of-distribution generalization where the overlap is hybrid, between a {\it classical} and a quantum state, 
single-qubit measurement suffice \cite{huang2024certifying}, provided 
each amplitude of the classical state can be computed efficiently. 
When the classical state is represented as an MPS, each amplitude requires $\mathcal O(N)$ operations, 
where $N$ is the number of qubits. 
Alternatively, each state $\rho(\bs x_n)$ can be mapped into a {\it shallow} quantum circuit \cite{cramer2010efficient,rudolph2023synergistic}, which can be implemented in a quantum device without exponential overhead.

Given the semi-analytic structure of $\Lambda$, we can also express the variance as
\begin{align}
	{\rm Var}(\Lambda)_{\rho(\bs x)} = \sum_{nm}&\alpha_n\alpha_my_ny_m\Big(\nonumber
		\Tr[\rho(\bs x_n)\rho(\bs x_m)\rho(\bs x)]-\\& \Tr[\rho(\bs x)\rho(\bs x_n)]\Tr[\rho(\bs x)\rho(\bs x_m)]\Big),
\end{align}
which can be classically estimated efficiently for the in-distribution case ($\bs x \in\mathcal S$),
allowing the learner to infer the 
number of measurements to be performed in the quantum device to achieve the desired precision. 
From H\"older's inequality, we get the bound ${\rm Var}(\Lambda) \leq (\sum_n \alpha_n)^2 \leq (CN_S)^2$,
where $N_S$ is the number of support vectors and $C$ is the regularization parameter 
-- typically $C=1$, see Appendix~\ref{sec:svm}. Clearly $N_S \leq M$, 
where $M$ is the number of training data. 
Therefore, the number of measurement shots to be performed scales at most quadratically in 
the number of training data. We will numerically show, using the generalization bounds from 
\cite{banchi2021generalization,banchi2024statistical}, that, for the in-distribution case, the number of training data 
to reach a desired accuracy scales linearly with the number of qubits. When this is 
the case, the overall complexity of our method scales polynomially with the number of qubits. 
When paired with \cite{huang2024certifying}, the whole measurement will require a polynomial number of 
single-qubit measurements. 

\begin{figure*}[t!]
	\centering
	\includegraphics[width=0.32\textwidth]{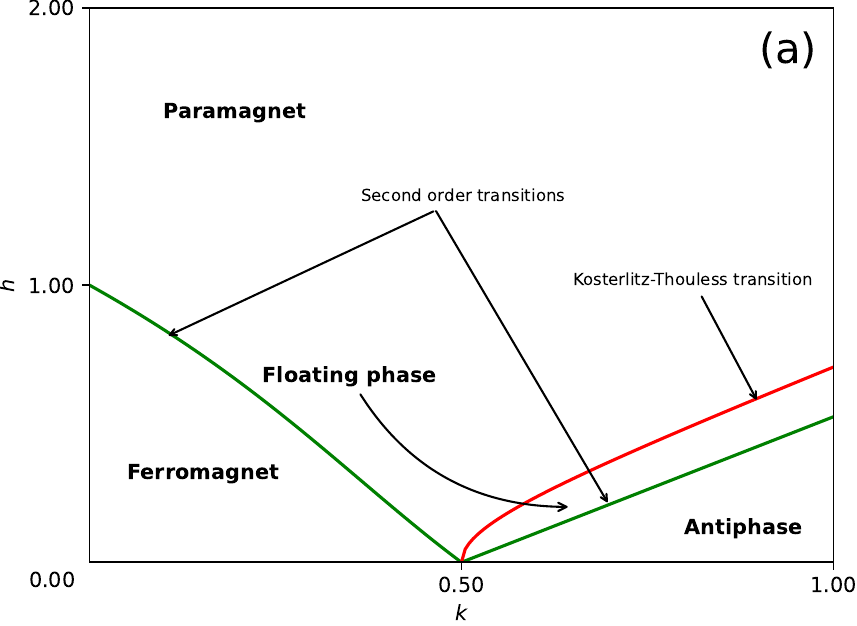}
	\includegraphics[width=0.32\textwidth]{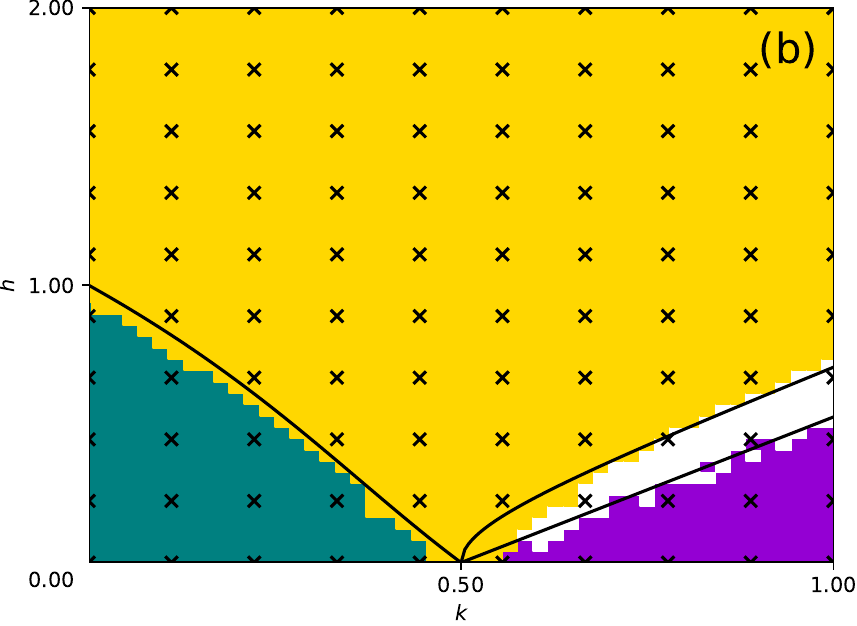}
	\includegraphics[width=0.32\textwidth]{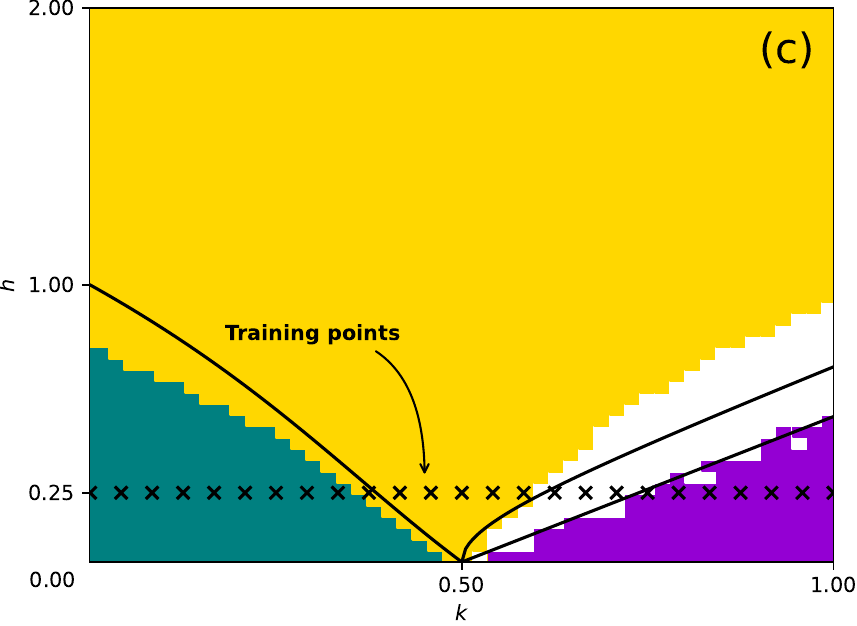}
	\caption{(a) Expected phase diagram of the ANNNI model in the thermodynamic limit. 
		(b-c) Predicted phase diagram for the ANNNI Hamiltonian \eqref{eq:annni} with $N=50$ spins, where the 
		ground state is simulated using tensor network methods with a small bond dimension (20). The training points 
		are marked with a black cross and the phases and phase transition lines are the same of Fig.~\ref{fig:annni}.
		In (b) the training point are uniformly scattered, while in (c) they are located on the line $h=0.25$.
	}
	\label{fig:annni}
\end{figure*}

\prlsection{Handling symmetries}
Symmetries may introduce degeneracies in the ground state. Consider 
the simplest case with two degenerate ground states labelled by $\ket{\psi_\pm(\bs x)}$.
For instance, if $H(\bs x)$ commutes with the parity operator, as in one of our
numerical results, $\pm$ may label the parity, and $\ket{\psi_\pm(\bs x)}$ will be
orthogonal, having two different parities. 
If the symmetry is not enforced by the diagonalization procedure, the ground state 
appears as a superposition 
\begin{equation}
	\ket{\psi_c(\bs x)}= \alpha_+ \ket{\psi_+(\bs x)} + \alpha_- \ket{\psi_-(\bs x)},
	\label{eq:parity superp}
\end{equation}
with possibly random coefficients $\alpha_\pm$. Those coefficients
represent a problem for kernel methods, as they introduce arbitrariness in the 
kernel entries \eqref{eq:kernel}, which might become random. 
To avoid this, during training we compute $\ket{\psi_\pm(\bs x)}$, e.g.,~by 
implementing the symmetry in the tensor network \cite{singh2010tensor}
and insert both states into the training set.
During test on the other hand, we assume we have less control and we do not implement 
any of those strategies. Therefore, each test ground state may appear as a possibly random 
combination of different states. This will be the case in ground states created by quantum 
devices, unless symmetries are explicitly enforced \cite{nguyen2022theory}. 
Nonetheless, at test the model should have learnt to 
separately classify the states with different parity, and so it should now be able to classify 
even states of the form \eqref{eq:parity superp}.

\prlsection{Numerics, ANNNI model}
As a first example, we focus on the 
one-dimensional Axial Next-Nearest-Neighbour Ising (ANNNI) model
\cite{elliott1961phenomenological,fisher1980infinitely,arizmendiPhaseDiagramANNNI1991,fumani2021quantum}, 
which describes $N$ spin-$\frac{1}{2}$ particles,  
with the Hamiltonian 
\begin{equation}
	H = -\sum_{j=1}^{N-1}\sigma_j^{\pauli x}\sigma_{j+1}^{\pauli x} + k \sum_{j=1}^{N-2}\sigma_j^{\pauli x}\sigma_{j+2}^{\pauli x} 
	-h \sum_{j=1}^N \sigma_j^{\pauli z},
	\label{eq:annni}
\end{equation}
where the two parameters $\bs x=(k,h)$ are both assumed to be positive. 
The above model displays a global $\mathbb Z_2$ symmetry, as $[H,P]=0$, where 
$P=\bigotimes_n \sigma_n^{\pauli z}$ is the parity operator. Therefore, energy eigenstates
can be degenerate and come as pairs with different parities.

The Hamiltonian \eqref{eq:annni} models ferromagnetic interactions between the $\pauli x-$components of nearest neighbouring spins (NN), 
and antiferromagnetic interactions (with relative strength $k$) between the same components of next-nearest neighbouring spins (NNN), 
together with the action of a magnetic field of strength $h$ along the $\pauli z$ direction.
For $k=0$ the model is exactly solvable using fermionization \cite{mbeng2020quantum}, 
while for $h=0$ the model is effectively classical, as the Hamiltonian is diagonal in the $\pauli x$ basis. 
For $k>0$, the system is frustrated, as the first term favours spin configurations with 
all spins aligned along the $\pauli x$ axis, while the second term favours configurations where 
next-nearest spins are anti-aligned.\footnote{Actually, the sign of the nearest neighbors interaction is not really relevant, 
as the Hamiltonian can be mapped into itself by changing the NN ferromagnetic interaction into 
an anti-ferromagnetic one, while rotating by an angle $\pi$ the spins on the even sites 
by a $\mathbb{Z}_2$ transformation $\bigotimes_{n~{\rm even}}\sigma_n^{\pauli z}$, explicitly proving that 
the frustrated nature of the model does not depend on the sign of the NN interaction, as two spin 
configurations cannot display exact anti-ferromagnetic order between next-nearest neighbours either when nearest neighbours are aligned or when they are anti-aligned.}

The phase diagram of the above model, shown in Fig.~\ref{fig:annni}(a),
has been characterised with different methods 
\cite{arizmendiPhaseDiagramANNNI1991,beccaria2007evidence,chandra2007floating,fumani2021quantum,suzukiQuantumIsingPhases2013,duttaQuantumPhaseTransitions2015}, and the possibility of using
tensor network techniques to compute its ground state was analyzed in 
\cite{beccaria2007evidence,nagy2011exploring}. For small $h$ and small $k$, the 
nearest neighbour interaction dominates, and the model displays ferromagnetic order; 
for sufficiently large $h$ the model is in a paramagnetic phase, while for large $k$ 
and suitably small $h$ the next-nearest neighbor interaction dominates and the model 
displays the so-called antiphase (AP), where pairs of aligned spins alternate pointing in opposite direction. Between the AP and the paramagnetic phase 
the model displays a floating phase, where correlation functions $\langle\sigma_j^{\pauli x}\sigma_{j+l}^{\pauli x}\rangle-\langle\sigma_j^{\pauli x}\rangle^2$ decay algebraically rather 
than exponentially, and the Hamiltonian becomes gapless in the thermodynamic limit $N\to\infty$. 

Aside from these general properties, the phase diagrams obtained 
with different methods do not always agree \cite{de2017quantum,nemati2020comment}:
Indeed, while there is a general consensus over the different phases schematically depicted 
in Fig.~\ref{fig:annni}(a), some works split the paramagnetic 
phase into two different phases \cite{fumani2021quantum}, while others found 
the floating phase over an extended region. 
For the phase diagram of Fig.~\ref{fig:annni}, 
the transition lines between the different phases have been found using different perturbative 
and numerical methods to approximately be \cite{beccaria2007evidence,cea2024exploring}
\begin{subequations}
\begin{align}
	h_{\rm I}(k) &\simeq \frac{1-k}k\left(1-\sqrt{\frac{1-3k+4k^2}{1-k}}\right), \\
	h_{\rm KT}(k) &\simeq 1.05 \,\sqrt{(k-0.5)(k-0.1)}, \\
	h_{\rm AP}(k) & \simeq 1.05\;(k-0.5).
\end{align}
\label{eq:transition lines}
\end{subequations}

\begin{figure}[t]
	\begin{center}
	\begin{tikzpicture}[scale=1, transform shape]
	\node at (-0.4,-0.4) {\includegraphics[width=0.26\textwidth]{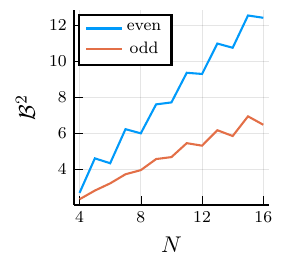}};
	\node at (4,0) {\includegraphics[width=0.22\textwidth]{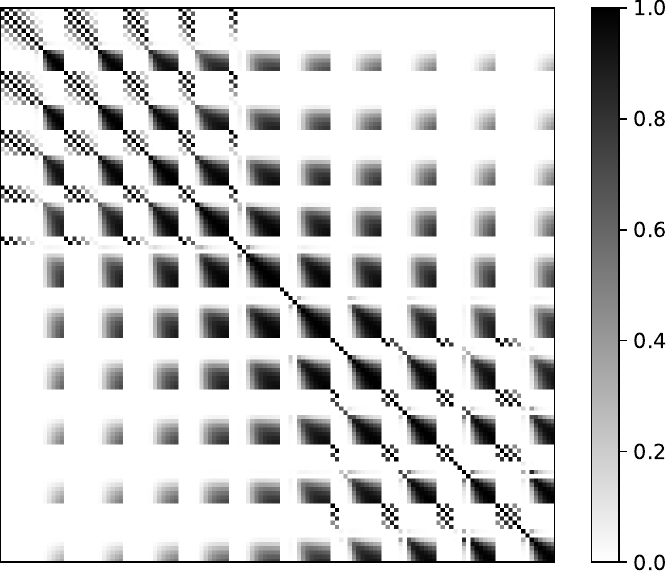}};
	\node at (1.05,-1.2) {(a)};
	\node at (5.05,1.4) {(b)};
	\end{tikzpicture}
	\end{center}
	\vspace{-7mm}
	\caption{(a) Scaling of the generalization error bound Eq.~\eqref{eq:bound}, for the ground state of the ANNNI model 
		with $h\in[0,2]$, $k\in[0,1]$, as a function of the number of qubits $N$, and 
		for the two different parity sectors (even and odd).
		(b) Kernel matrix entries for the training states shown in Fig.~\ref{fig:annni}. 
	}
	\label{fig:gen}
\end{figure}

Our numerical results for the ANNNI model are shown in Fig.~\ref{fig:annni} for a chain with $N=50$ spins. 
The ground states were obtained using the DMRG algorithm implemented in the \verb+quimb+ library \cite{gray2018quimb} .
In spite of the small bond dimension (20), the predicted phase diagram is in good agreement with the theoretical 
prediction, shown in Fig.~\ref{fig:annni}, with differences only near the phase transition points.
These differences might be due to finite size effects, as the phase diagram in Fig.~\ref{fig:annni} is expected 
in the thermodynamic limit $N\to\infty$, and due to the finite bond dimension, as close to the phase transition point 
the entanglement area law breaks down \cite{eisert2010colloquium} and a larger bond dimension might be needed.
This highlights a use case in which a classically learned observable could be measured in a shallow quantum circuit for ground states near the transition point in order to locate the phase boundary with greater precision than is possible purely classically.

In Fig.~\ref{fig:annni} we trained the model in two different ways, and then tested over the same parameter region. 
In panel (b), the training points are uniformly spaced (in-distribution), while in panel (c) they are chosen along a line 
that passes through all four different phases (out-of-distribution). Prediction in (b) is simpler, as the model has to interpolate 
over the training points, while in (c) it is more complex as the model has to extrapolate, resulting 
in a lower accuracy. Nonetheless, in both cases the model displays remarkable generalization abilities given 
the limited amount of data. 

The ability of the training algorithm to generalize, namely to predict the phase of ground states outside of the training set,
can be explained using the language of Ref.~\cite{banchi2021generalization,banchi2024statistical} when the training and test data belong to the same distribution. 
Indeed, the generalization error, namely the difference between an optimal classifier with full knowledge of the data distribution 
and an empirical classifier that only has access to $M$ samples from that distribution, is upper bounded by $\mathcal B/\sqrt M$ 
where $\mathcal B$ depends on the mutual information between the parameter space and the set of possible ground states. 
The in-distribution generalization error is small as long as 
$M\gg \mathcal B^2$. The values of $\mathcal B^2$, estimated by following Appendix \ref{sec:generr}, 
are shown in Fig.~\ref{fig:gen}(a), 
where we show that $\mathcal B^2$ increases at most linearly with the number of qubits. 
Therefore, in the worst case, at most a number of
training points that scales linearly with the number of qubits is sufficient to ensure a good generalization, in spite 
of the exponentially growing Hilbert space. 
In practice though, as discussed in \cite{banchi2024statistical}, because of the regularization term, the above bound might be loose 
and the model might be able to generalize even with less data. 
The good generalization capabilities are due to the ``clustering'' of the ground states within the different phases, which is 
evident from the kernel matrix entries shown in Fig.~\ref{fig:gen}(b).
In that figure, the ``tiled'' subregions are due to the degeneracy of the ground states, which can be labeled by the parity and 
are mutually orthogonal. 

When training and test data belong to different distributions, as in Fig.~\ref{fig:annni}(c), 
we need to employ out-of-distribution bounds, which have been discussed in the quantum case 
only for different settings \cite{caro2023out}.

\prlsection{Numerics, Haldane model}
As a second model, we focus on the  one-dimensional 
symmetry-protected topological 
spin model described by the Hamiltonian \cite{cong2019quantum}
\begin{equation}
	H = -\sum_{j=1}^{N-2}\sigma_j^{\pauli x}\sigma_{j+1}^{\pauli z}\sigma_{j+2}^{\pauli x} - k \sum_{j=1}^{N-1}\sigma_j^{\pauli z}\sigma_{j+1}^{\pauli z} 
	-h \sum_{j=1}^N \sigma_j^{\pauli z},
	\label{eq:Htopo}
\end{equation}
with parameters $\bs x= (h,k)$. 
When $k=0$ the model is exactly solvable via the Jordan-Wigner transformation \cite{mbeng2020quantum}. 
Such a Hamiltonian has a $\mathbb Z_2\times\mathbb Z_2$ symmetry 
generated by the operators $\hat X_{\rm even(odd)} = \prod_{i\in{\rm even(odd)}} \sigma_i^{\pauli z}$. 
The ground state of $H$ displays, as a function of $h$ and $k$, 
a paramagnetic phase, an antiferromagnetic phase and 
a $\mathbb Z_2\times\mathbb Z_2$ symmetry-protected topological (SPT) phase.
The latter can be detected by non-zero string order parameters 
$S_{ab} = \sigma_a^{\pauli x} \prod_{a<i<b}(\sigma_i^{\pauli z}) \sigma_b^{\pauli x}$ or with 
a quantum convolutional neural network circuit \cite{cong2019quantum}.

\begin{figure}[t]
	\centering
	\includegraphics[width=0.4\textwidth]{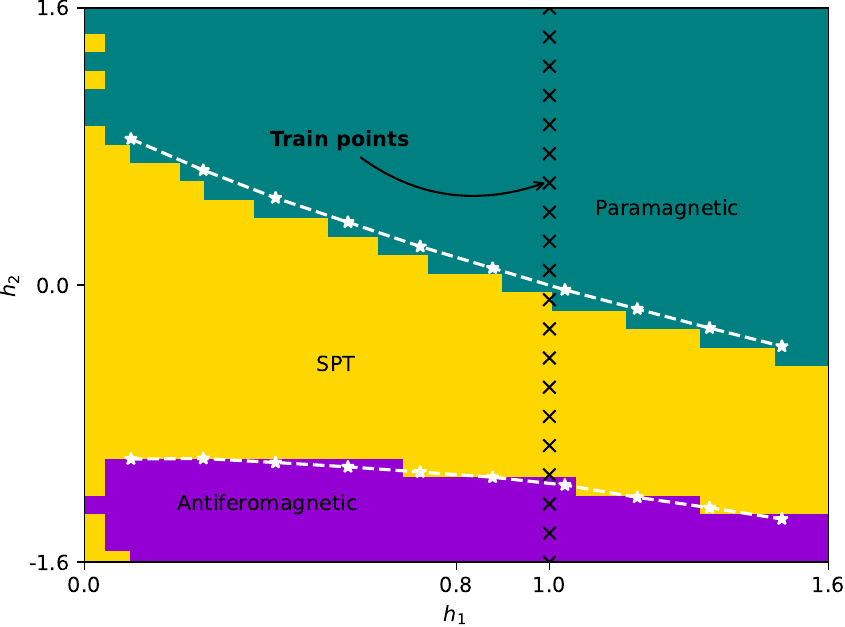}
	\caption{
	  Predicted phase diagram for the Hamiltonian \eqref{eq:Htopo} with $N=41$ spins, where the 
		ground state is simulated using tensor network methods with a bond dimension 150. The training points, 
		marked with black crosses, are located along the line $h=1$. 
		The phase transition lines (dashed white) are estimated by interpolating the numerically observed 
		points (white asterisks) from Ref.~\cite{cong2019quantum}. 
	}
	\label{fig:topological}
\end{figure}

To train the model, we fix $N=41$ and provide just the known information from Refs.~\cite{cong2019quantum,sone2024quantum} that 
the model belongs to the SPT phase for $h=1$ and $-1.15 \leq k\leq 0$, to the paramagnetic phase 
for $k>0$ and to the antiferromagnetic phase for $k<-1.15$. The 50 training points are displayed with 
black crosses in Fig.~\ref{fig:topological}. 
We then test the model over the entire phase space shown in Fig.~\ref{fig:topological}, discretized 
as a $30\times30$ grid. For comparison, we also draw the phase transition lines obtained in \cite{cong2019quantum}. 
From the results shown in Fig.~\ref{fig:topological} we see that the SVM classifier is able to accurately recognise 
all three phases, generalizing the given information from the states along the $h=1$ line. 
Imperfections around the phase transition lines are possibly due to finite size effects or due to the small bond-dimension 
in the tensor network. 
Misclassifications around the $h=0$ line are comparable to those obtained with other methods \cite{cong2019quantum,sone2024quantum} 
and may be exacerbated by the fact that the states around $h=0$ might be qualitatively quite different  from the training 
states along $h=1$.

\prlsection{Conclusions}
We studied the classification of quantum phases of matter using a combination of classical and quantum techniques. 
We studied how to use knowledge from part of a phase diagram to populate other, possibly unknown,
parts of the phase diagram. The result of our learning approach is a set of observables that can be measured 
in quantum devices, with a number of measurements that scale polynomially in the system size. 
We studied different models with rather rich phase diagrams and showed that our classifier is able to 
both interpolate and extrapolate the information from the training set to predict the quantum phase for 
new sets of parameters. 

\prlsection{Acknowledgments}
The authors acknowledge financial support from: PNRR Ministero Università e Ricerca Project 
No. PE0000023-NQSTI funded by European Union-Next-Generation EU (M.K., A.C. and L.B:); 
Prin 2022 - DD N.~104 del 2/2/2022, entitled ``understanding the LEarning process of QUantum
Neural networks (LeQun)'', proposal code 2022WHZ5XH, CUP B53D23009530006 (L.B.); Prin 2022 - DD N.~104 del 2/2/2022, proposal code 2022SJCKAH, CUP B53D23005250006 (A.C.);
U.S. Department of Energy, Office of Science, National Quantum Information Science Research
Centers, Superconducting Quantum Materials and Systems Center (SQMS) under the Contract
No. DE-AC02-07CH11359 (J.P., A.C. and L.B.).

\appendix

\section{Quantum-enhanced support vector machines}\label{sec:svm}

Support Vector Machines \cite{cristianini2000introduction}
are a family of machine learning models based on 
decision hyperplanes on a Hilbert space, called \emph{feature space}. 
Training consists in finding such a hyperplane, while classification of 
a new input is done by checking on which side of the plane the input belongs. 
Since multiclass classification problems are broken down into a cascade of binary decisions,
we will focus on binary classification problems with $Y=\{+1,-1\}$.
Given a fixed map $\bs x\mapsto \rho(\bs x)$, 
embedding a classical variable $\bs x$ into the quantum (feature) space,
the linear decision hyperplanes in the feature space yields a non-linear 
hypersurface in the input space,  defined as
\begin{equation}
	f(\bs x)= \Tr[W\rho(\bs x)] +b=0.
\label{eq:hypersurface}
\end{equation}
The above equation describes a hyperplane in the feature space, with parameters
encoded into the ``observable'' $W$ 
and the shift $b$. It can be shown that the optimization of those parameters 
during training can be 
cast into a ``dual'' formulation that can be solved with
open-source libraries \cite{chang2011libsvm} with polynomial complexity in 
the number of training pairs $M$. 
The resulting  convex optimization problem is described by 
\begin{equation}
	\argmax_{\{0\leq \alpha_i < C\} }\left[\sum_{i=1}^{M}\alpha _{i}-\frac{1}{2}\sum _{i,j=1}^{M} 
y_{i}y_{j}\alpha _{i}\alpha _{j}k(\bs x_i,\bs x_j)\right],
\label{eq: dual opt}
\end{equation}
with constraints ${\sum }_{i=1}^{M}\alpha _{i}y_{i}=0$  and a hyper-parameter $C$.
In the above equation the kernel is defined as in Eq.~\eqref{eq:kernel}, 
and $(\bs x_i,y_i)$ for $i=1,\dots,M$ define the $M$ training pairs with inputs $\bs x_i$ 
and known label $y_i$.
Training consists in finding the coefficients $\alpha_i$. Once those are obtained, 
we can get the ``observable'' and shift from Eq.~\eqref{eq:hypersurface} analytically as
\begin{align}
	W&=\sum _{i=1}^{M}y_{i}\alpha _{i}\rho(\bs x_i), \\
	b&=\sum _{i=1}^{M}\left(\frac{y_i -
			\sum_{j=1}^M\alpha_jy_jk(\bs x_i,\bs x_j)}M\right),
\end{align}
as well, as the predicted class $y$ of a new test $\bs x$, via 
the decision function  \eqref{eq:hypersurface}
\begin{align}
\label{eq: f}
y = {\rm sign}[ f(\bs x)]& ={\rm sign}\left[b+ \sum _{i=1}^{M}\alpha _{i}y_{i}k\left(\bs x_{i},\bs x\right)\right]. 
\end{align}

We conclude this appendix by clarifying the differences between the classification of 
quantum and classical data. 
In many applications of quantum kernel methods 
\cite{schuld2021supervised} we are interested in classifying complex classical data $\bs x$, 
e.g.,~images, by first embedding those images into a quantum state $\ket{\psi(\bs x)}$. 
However, there is no general rule to define such an embedding circuit, which can then 
be trained as well \cite{gentinetta2023quantum,montalbano2024quantum}. 
On the other hand, in our case, the classical parameters $\bs x$ are much 
less complex (often described by two real numbers rather than by the many pixels of an image), 
and the encoding map $\bs x\mapsto{\rho(\bs x)}$ is fixed by nature, being the map 
that associates to any $\bs x$ the ground state of $H(\bs x)$.  
For classical data, the purpose of the kernel is to define a non-linear map into a much 
larger Hilbert space, called the feature space, where complex classical data are expected to be easier to discriminate.
In our case the ``feature'' space is the physical Hilbert space of density matrices of quantum many-particle systems.

\section{Comparison with previous studies} \label{sec:comparison}

In this section we clarify the main differences between our approach and Refs.~\cite{sancho2022quantum,wu2023quantum}. 
First of all, we point out that the main ones are due to the hybrid nature of our method, which is physically motivated (we train with 
what we know, and test with what we don't), and algorithmically efficient, when paired with 
\cite{huang2024certifying}. 
Indeed, the algorithm from \cite{huang2024certifying} is efficient in this hybrid regime, where we have an effective classical description of one of the two states. 
Clearly though, in order to verify the accuracy of the learnt model, we had to test it for parameter regions where 
we had full knowledge. In what follows we highlight other major differences. 

Compared to Ref.~\cite{sancho2022quantum}, 
in our work we use the fidelity squared rather than the fidelity to define the kernel. This seemingly 
minor change provides us a physical interpretation of the decision hyperplane as a physical observable
quantity, which is lacking for the fidelity kernel \cite{schuld2021supervised}. 
Moreover, the authors of Ref.~\cite{sancho2022quantum} applied their algorithm to the Ising model, where quantum phases are easy to learn \cite{banchi2021generalization,banchi2024statistical} and the fidelity can be computed analytically, while in this work we apply our method to more challenging phase diagrams. 
Compared to Ref.~\cite{wu2023quantum}, we use a different cost function, which allows us to use 
standard algorithms \cite{chang2011libsvm} for processing the kernel entries, and we use 
classical tensor network methods during training, while \cite{wu2023quantum} uses parametric quantum circuits 
and a quantum variational eigensolver to approximate the ground state (which requires a quantum device) \cite{bharti2022noisy}. 
It is unknown whether the latter 
can provide an advantage over classical methods \cite{cerezo2023does},  e.g.,~based on tensor networks,
while it certainly adds the extra issue of the measurement cost in estimating expectation values.
Ref.~\cite{wu2023quantum} also conjectures the \textit{worst case} difficulty of classical training, whilst we focus on specific, physically relevant scenarios in which classical training can be demonstrated to be efficient.
Compared to both Refs.~\cite{sancho2022quantum,wu2023quantum} we also explicitly consider degeneracies that,
without a careful treatment, might make kernel entries a random number, thus affecting the ability of the model 
to learn. 

\section{Generalization error bounds}\label{sec:generr}
Following~\cite{banchi2021generalization,banchi2024statistical}, since ground states are pure, the bound 
quantity $\mathcal B$ can be estimated using $m$ samples (possibly different from the number 
of training samples $M$) as 
\begin{align}
	\mathcal B &\simeq \Tr\sqrt{\mathcal K}, & \mathcal K_{jk} &= \bra{\psi(\bs x_j)}\psi(\bs x_k)\rangle / m,
	\label{eq:bound}
\end{align}
for $i,j=1,\dots,m$;
the error in approximating $\mathcal B$ with $m$ samples goes at most as $\mathcal O(m^{-1/2})$ \cite{banchi2024statistical}. 
Note that the above kernel matrix differs from \eqref{eq:kernel}, as $K_{jk}=|\mathcal K_{jk}|^2/m$. 
We estimated the values of $\mathcal B$ shown in Fig.~\ref{fig:gen} by
using $m=1000$ random uniform samples of $h\in[0,2]$ and $k\in[0,1]$, end then checked that $\mathcal B$ 
converged, namely that by slightly increasing the number of samples the outcomes were basically the same. The final value
is estimated by averaging over 10 repetitions to average out the (small) fluctuations.

\end{document}